\documentclass[copyright,creativecommons]{eptcs}

\usepackage{breakurl}        
\usepackage{epsf}
\usepackage{graphicx}

\title{Modelling and Verification of Multiple UAV Mission Using SMV}
\author{Gopinadh Sirigineedi\thanks{PhD Student., DoIS, Cranfield University.},
 \ Antonios Tsourdos\thanks{Professor and Head of the Autonomous Systems Group., DoIS, Cranfield University.},
  Rafa{\l} {\.Z}bikowski\thanks{Professor., DoIS, Cranfield University.},
 \ and \ Brian~A. White\thanks{Professor., DoIS, Cranfield University.}\\
 \\ 
 {\normalsize\itshape
 Department of Informatics and Sensors,}\\
 {\normalsize\itshape
 Cranfield University, Shrivenham,}\\
 {\normalsize\itshape
 Swindon SN6 8LA, United Kingdom.}}

\begin{document}
\maketitle

\begin{abstract}
  Model checking has been used to verify the correctness of digital
  circuits, security protocols, communication protocols, as they can
  be modelled by means of finite state transition model. However,
  modelling the behaviour of hybrid systems like UAVs in a Kripke
  model is challenging. This work is aimed at capturing the behaviour
  of an UAV performing cooperative search mission into a Kripke model,
  so as to verify it against the temporal properties expressed in
  Computational Tree Logic (CTL). SMV model checker is used for the
  purpose of model checking.
\end{abstract}

\section{Introduction}
Increase in computational power, improvements in control techniques and other technological advances have led to increased focus on cooperative control of multiple agents in recent years. Cooperating multi-agent systems find application in large number of areas - mobile robots, micro satellite clusters, unmanned aerial vehicles (UAVs), automated highway systems and internet agents. Cooperating multiple-agents offer a large number of benefits such as: increase in the success rate of mission, large area coverage by improvements in latency and information gathering, increase in the computational power offered by distributed computing and graceful degradation in performance.
\par
Cooperative UAV control problems that have received recent attention include cooperative formation \cite{Park}, cooperative task allocation, cooperative path planning, cooperative search and many others. Cooperative search problems have applications in a number of military and civilian applications, such as surveillance and reconnaissance operations, search and rescue; hazard monitoring, battle damage assessment, agricultural coverage tasks and security patrols \cite{Yang}.
\par
Due to the mission critical nature of UAV systems, it is highly important to ensure the correctness of these systems and check if the system meets the design requirements.  The failure of control software of the Arian-5 rocket and the Mars rover are hard remainders of what can happen when systems don't perform as per specifications. Verification is the process of verifying the correctness of the system and whether it satisfies the specifications. The common verification processes are \textit{simulation, testing, deductive verification}, and \textit{model checking} \cite{ModelChecking}. Simulation is performed on the abstract model of the system, where as testing is done on the actual system. Simulation and testing involve giving certain inputs and checking whether the outputs are as expected. These methods are cost effective way to find the errors. However, checking all of the possible interactions and possible faults is almost impossible using simulation and testing. They only ensure that the system works for the inputs they are tested for. Testing will demonstrate the presence of bugs, but will not demonstrate the absence of bugs. Even if the system passes all the testing, we can't claim that the system is completely free from errors, as no amount of testing is exhaustive enough.

\par
The mission planning software of multiple UAV systems involves concurrency as it deals with multiple UAVs. It is also reactive, as it constantly interacts with the environment in which the UAVs operate.  It is impossible to completely verify such a software system using traditional testing. In addition, concurrency bugs are one of the most difficult ones to test in a traditional way. Moreover, autonomous systems operate in harsh and unpredictable environments and it is difficult to predict before hand the kind of situations that may arise during the mission to carry out testing \cite{Hinchey} \cite{Gat} \cite{ChristopherRouff}. Hence, there is a need for formal verification methods, like deductive verification and model checking, which can clarify with high degree of certainty that the system meets its requirements. NASA has been working on developing formal verification techniques for their intelligent autonomous system involving multiple rovers or satellites \cite{Brat} \cite{Pecheur}.

\par
Deductive verification is proof-based. It refers to axioms and proof rules to prove the correctness of the system. System description is made in some formal language and leads to a set $\Gamma$ of appropriate formulas in an appropriate logic. The set $\Gamma$ constitutes a formal logical inference system for deduction. A system specification is an another formula $\varphi$ of a chosen logic. The verification consists of finding a proof within the given formal logical system, which would demonstrate that the specification formula $\varphi$ is inferred from the axioms and inference rules of the formal system $\Gamma$, \textit{ie} $\Gamma\models\varphi$. The formal system $\Gamma$ is assumed to be sound and complete. Deductive verification is well recognized in computer scientists and has significantly influenced the area of software development. However, deductive verification is time-consuming and can be performed only by experts in logic and mathematics.
\par
Model checking, as the name suggests, is model-based. It is an automatic technique for verifying finite state systems. It involves developing a simplified model which captures the essential features of the systems. The specifications which are to be verified on the system are specified, usually in terms of logical statements. Then a model checker, a software tool, systematically examines all the system scenarios to check whether the system satisfies the specifications \cite{Clarke}.
\par
In \cite{Jayaraman1} SPIN model checker has been used to verify the safety properties of multi-robot system expressed in Linear Temporal Logic (LTL). In \cite{MultirobotPlanningATimedAutomataApproach} timed automata framework has been used to model the robots and \textsf{Uppaal} model checker has been used to verify the properties of multiple-robot system expressed in Computational Tree Logic (CTL). This paper presents formal modelling of multiple-UAV mission by means of Kripke model and verification of some of the mission properties expressed in CTL. Kripke model offers benefits of using graph theoretic approaches to analyze the system model. SMV model checker is used for verifying the properties, as it is one of the most popular model checkers that supports CTL. In our previous work \cite{TowardsVerifiableApproach}, we reported Kripke model of the behaviour of a single UAV performing a search and verification of its properties expressed in CTL.
\par
The rest of the paper is organized as follows: Section-2 gives an overview of model checking technique. The multiple UAV mission and single UAV behaviour performing the search are discussed in Section-3. Verification of the UAV behaviour using SMV model checker is presented in Section-4.

\section{Overview of Model Checking }\label{sec:modelchecking}
\par
Model checking is a technique to verify finite state machine abstraction of the system. The system is represented by finite state model $M$ and a temporal logic formula $\phi$ expressing some desired specification. A model checker is used to check $M$ against the specification $\phi$. The model checkers outputs either true, if $M$ satisfies $\phi$ \textit{i.e} $M \models \phi$, or a counter example, if it does not. These different steps of model checking are discussed in detail below.

\subsection{Model}
The first step in model checking is to construct a \textit{formal model} of the system. As model checking can be performed only on finite state systems, the system should be represented as a finite state transition diagram. We are primarily concerned with reactive systems like UAV systems and their behaviour over time. Reactive systems are systems which maintain constant interaction with the environment in which they operate. The family of reactive systems include many classes of programs whose correct and reliable construction is particularly challenging, including concurrent programs, embedded and process control programs, and operating systems. Typical examples of such systems are air traffic control systems, operating systems, and perpetual ongoing processes such as a nuclear reactors.
\par
Reactive systems need to interact with the environment frequently and often do not terminate. Therefore, they can't be adequately modelled by input-output behaviour. Reactive systems can be modelled by capturing the features by means of \textit{state}. A state is an instantaneous description of the system that captures the variables at a particular instant of time. The change from one state to the other as a result of some action determines the \textit{transition} of the system. A \textit{computation} is an infinite sequence of states where each state is obtained from the previous state by some transition.
\par
\textit{Kripke structure} or \textit{Kripke model} is a type of state transition graph to capture this intuition about the behaviour of reactive systems. Kripke structure consists of a set of states, a set of transitions between the states, and a function that labels each state with a set of properties that are true in that state. Paths in Kripke structure model computations of the system. Although the model of the system is abstract and simple, it should be expressive enough to capture the aspects of temporal behaviour for reasoning about the system. Formal representation of Kripke structure is given below.
\par
\begin{figure}[thpb]
\centering
\includegraphics[height=50mm]{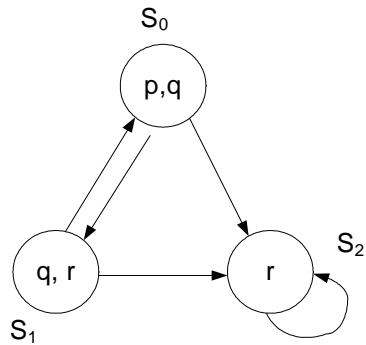}
\caption{Kripke model representation of a finite state system} \label{fig:KripkeEx}
\end{figure}

A Kripke model is represented by a triplet $M = (S,R,L)$ over a set of atomic propositions $AP$ \cite{ModelChecking}. A concise representation of a Kripke model whose nodes are states is shown in Figure \ref{fig:KripkeEx}.
\begin{enumerate}
  \item $S$ is a finite set of states.
  \item $R \subseteq S \times S$ is the transition relation.
  \item $L: S \rightarrow 2^{AP}$ is a function that labels each state with the set of atomic propositions true in that state.
\end{enumerate}

\subsection{Specification}
The properties of the system are usually specified in temporal logic. Temporal logic is a formalism for describing sequences of transitions between states in a reactive system. It is used for specifying and verifying the correctness of digital circuits and computer programs. In classical logic, such as predicate logic, the truth value of a statement is independent of time, whereas in temporal logic the truth value changes dynamically with time. As we are trying to capture the  behaviour of multiple UAV group over time, temporal logic suits the purpose of specification language for specifying requirements of the system. There are two fundamental representation types of temporal logic: CTL and LTL. The distinction is how they handle the underlying computation tree.
\par
CTL considers branching of time and allows future paths at any given point of time. The temporal operators quantify over the paths that are possible from a given state. The CTL operators are AX, EX, AG, EG, AU, EU, AF and EF. These operators are pairwise operators. The first of the pair is either A or E. A represents `along all paths' and E represents `there exists at least one path'. The second one of the pair is X, F, G, or U meaning `next state', `some future state', `all future states(globally)' and `until' respectively. CTL has the following syntax given in Backus Naur form:
\begin{equation}\label{th}
    \phi:= \bot|\top|p|(\phi)|(\neg\phi)|(\phi\wedge\phi)|(\phi\vee\phi)|\phi\rightarrow\phi|AX\phi|EX\phi|\\
           A[\phi\cup\phi]|E[\phi\cup\phi]|AG\phi|EG\phi|AF\phi|EF\phi,
\end{equation}
where $p$ ranges over atomic formulas.

\subsection{Verification}
Many automatic model checkers are available, e.g. SMV \cite{McMillanSMV}, SPIN\cite{ModelCheckerSPIN}, KRONOS, HYTECH, NuSMV. SMV model checker has been developed by McMillan in the 90s. It uses SMV language for description of the system model. It accepts temporal specifications expressed either in CTL or LTL. SMV has been used for model checking digital circuits \cite{Kotmanova}, security protocols \cite{GigabitEthernet}, embedded systems \cite{PowerManagement} and web applications \cite{WebApplicationsSMV}.

\section{Mission planning for Multiple UAVs}
In recent years there has been a growing interest in employing UAV teams to cooperatively search a given area. The operations of such groups include reconnaissance, surveillance, battle damage assessment, fire monitoring and chemical cloud detection. UAVs are suitable for these operations as they are too dangerous for human pilots. Some of the tasks such as monitoring forest fire propagation and mapping chemical cloud propagation can only be carried out by a group of UAVs and can't be carried out by a single UAV \cite{Beard}, \cite{EvolvingBehav}, \cite{Jin}, \cite{Yang}.
\par
\begin{figure}
\centering
\includegraphics[width=9cm]{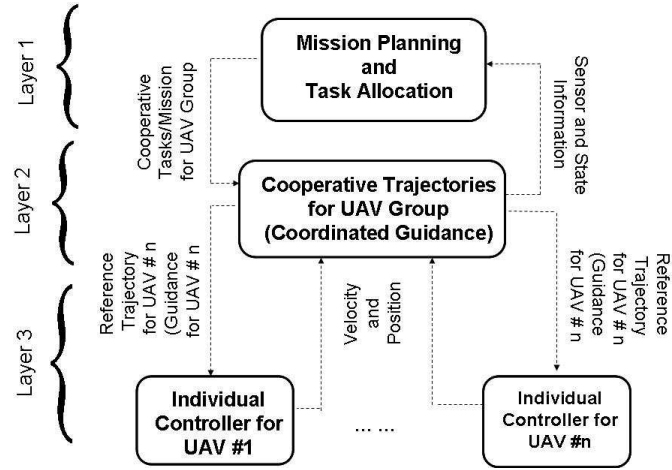}
\caption{Architecture for multiple-UAV cooperative control} \label{fig:architecture}
\end{figure}

Hierarchical approach is used extensively for cooperative control because of its simplicity and ease of design \cite{Beard} \cite{Boskovic} \cite{Coordinated_target_assignment}. Each layer has different functionalities. This is done to simplify design and to deal with different aspects of the systems in separate layers. Partitioning may result in sub-optimal solution, but as each of these layers have different bandwidths dealing them separately can be justified. The decision making layer operates at very low rate, whereas the path planning layer operates at moderate rate and auto-pilot operates at high rate. The control architecture for multiple UAV mission is shown in Figure \ref{fig:architecture}. The top layer performs as decision making layer taking inputs regarding the state of the UAVs from the middle layer. The middle layer is the path planning and guidance layer. The bottom layer consists of controllers for each UAV.

\subsection{Cooperative UAV Search}
This research is aimed at developing a verifiable multiple UAV mission for cooperatively searching a given area. The mission is to search a given bounded area for static targets and threats using a group of UAVs for the purpose of environmental monitoring. No or little information is available about the area being searched. The UAVs must cooperatively search the environment and mark the positions of targets and threats. Each UAV has a communication link to broadcast the findings and to receive updates from other UAVs of the group. Each UAV has sensors to monitor the environment for the presence of threats and targets. It is assumed that inter-UAV communications are instantaneous, noiseless and have unbounded communication range. It is assumed that each UAV has sufficient processing power for path planning and enough storage to store the global picture of the area being searched. The velocity of the UAVs is 20 m/s and the minimum turn radius is 25 m.

\par
Each UAV stores on board a model of the environment in form of a ``search map''. The positions of targets, positions of threats identified by the UAVs in the group and decisions of the other UAVs in the group are stored in this search map. This search map is constantly updated to reflect new information gathered and changes in the state of the UAVs. Sharing information is essential to ensure cooperation for decentralized control approach \cite{CoordinatedMultiRobotExploration}, \cite{Yang}. As no centralized control is present, sharing of information among the UAVs helps in achieving cooperation. The environment being searched is divided into square cells. Based on the information in search map each UAV will identify an adjacent cell free from threats and other UAVs. The path planning layer generates a path starting from its present position, to the selected neighbouring cell with selected intial and destination headings.

\par
\begin{figure}
\centering
\includegraphics[width=6cm]{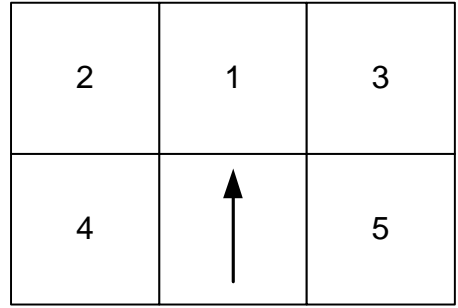}
\caption{Neighbouring cells marked for decision making} \label{fig:NeighbouringCells}
\end{figure}

The area to be searched is a square area of 2000X2000 metres. The area is discretized into square cells of 100x100 metres. Discretizing the area into cells helps in reducing the state space and also to visualize the state of location of the UAV. Each cell is identified by coordinates of the centre of cell. The UAVs move through the search area by selecting one of the neighbouring cells. The neighbouring five cells around the cell, in which UAV is flying, are marked as shown in Figure \ref{fig:NeighbouringCells} for the purpose of identification. The cell with arrow corresponds to the cell in which the UAV is flying at the time of decision making and the direction of arrow indicates the current heading of the UAV. Cell marked with 1 corresponds to the cell just ahead in the direction of current heading. When cell1 is free, the UAV moves into it. If cell1 contain a threat or it is already chosen by other UAV in the group, the UAV moves into cell3. The order of preference in selecting a neighbouring cell in decreasing order is - cell1, cell3, cell5, cell2 and cell4. When the UAV reaches the north-most cell it moves into cell5. When the UAV reaches the south-most cell it moves into cell4. The UAV flies in a path which connects the centre of present cell and the centre of the chosen neighbouring cell. The initial heading and destination headings of the path are either 90$^\circ$ or 270$^\circ$.  This is done to discretize the heading of the UAV to just two values in order to capture the UAV heading in finite state transition model. The heading is measured with respect to the east. Heading of 90$^\circ$ corresponds to the UAV flying north. The destination heading of the path is same as the initial heading if the UAV selects cell1 or cell2 or cell3. If not, the destination heading is opposite direction to the initial heading. Each UAV in the group repeats this behaviour of selecting a neighbouring free cell.

\begin{figure}[thpb]
\centering
\includegraphics[width=10cm]{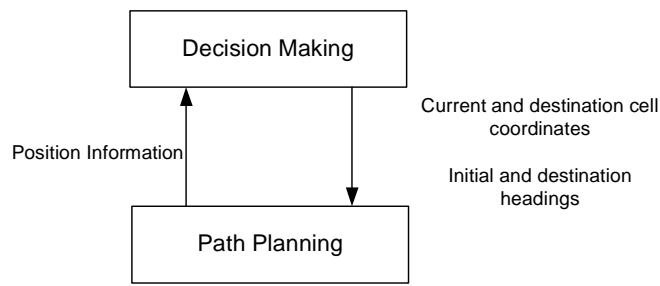}
\caption{Information exchange between decision making layer and path planning layer} \label{fig:SingleUAVLayeredArchitecture}
\end{figure}

The decision making layer passes the information of the current and destination cells, current and destination headings to the path planning layer as shown in Figure \ref{fig:SingleUAVLayeredArchitecture}. The path planning layer takes inputs from the decision making layer and generates a flyable path from the current cell to the destination cell with specified starting heading and destination heading. The flight dynamics of the UAV are not taken into account. It is assumed that the flight controller present on the UAV will take inputs from the path planning and guidance layer and follows the generated path.

\begin{figure}[thpb]
\centering
\includegraphics[width=9cm]{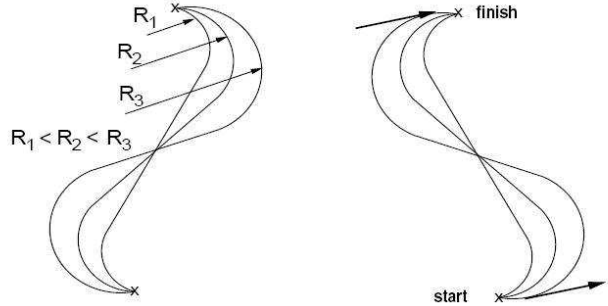}
\caption{2D Dubins paths for different turning radii and heading angles} \label{fig:DubinsPaths}
\end{figure}

Dubins paths are used to generate path between the way points identified by the decision making layer. The decision making layer passes the inputs to the path planning layer which then produces a Dubins path for the initial and destination poses. Dubins path produces the shortest path between two points by concatenation of circular arcs and their connecting tangents \cite{Dubins} \cite{Classification_of_dubins_sets}. A straight line is the shortest path between two points. However, for UAVs with constraints on initial heading and final heading and minimum turning radius, the shortest path is given by concatenation of circular arcs and straight line \cite{Dubins}. Four Dubins curve types \textit{LRL}, \textit{RSR}, \textit{RSL}, \textit{LSR} have been used. \textit{L}, \textit{S}, \textit{R} denote turning left, straight line and turning right respectively. It is assumed that the UAVs fly with a constant altitude. Hence, 2D Dubins paths are used for path planning. 2D Dubins paths for different turning radii and different heading angles are shown in Figure \ref{fig:DubinsPaths}.

\begin{figure}[htpb]
\centering
\includegraphics[width=15cm]{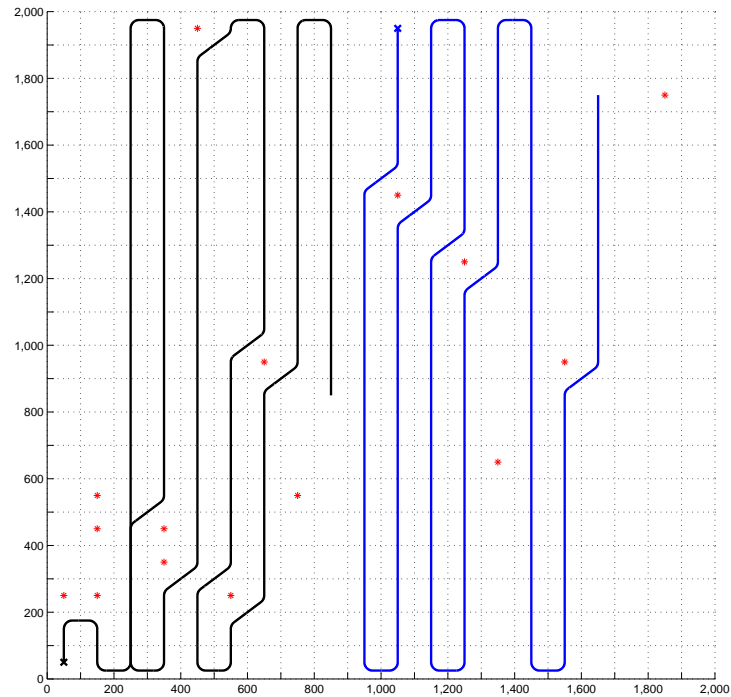}
\caption{Simulation result of multiple UAV mission} \label{fig:MultipleUAVSearch}
\end{figure}

The search strategy discussed above is implemented in \texttt{MATLAB}. The trajectories of two UAVs in the group for a flying time of 600 seconds is shown in Figure \ref{fig:MultipleUAVSearch}. The red dots indicate the cells with threats.

\section{Model Checking UAV Misssion}
\subsection{Kripke Model of UAV Mission}
The behaviour of an UAV in the group performing the mission has been captured in Kripke model. CTL has been used to specify the properties because of its expressiveness and ease of modelling in Kripke model. As SMV model checker is a popular model checker for checking CTL properties is has been chosen for our work. SMV language is used for the description of the corresponding Kripke model.

\par
The state of the UAV is captured by the current values of the programs variables. The state of the position of the UAV is captured by means of the coordinates of the centre of the cell. State variables \texttt{current\_cell} and \texttt{destination\_cell} capture the present and destination cell coordinates respectively. The initial values for \texttt{current\_cell}, \texttt{initial\_heading} are assigned using the keyword \texttt{init} as shown in Figure \ref{fig:SMVCodeInit}.
\begin{figure}[hbpt]
\centering
\begin{verbatim}
 init(current_cell) := [10 , 10];
 init(initial_heading):=90;
\end{verbatim}
\caption{SMV code showing assignment of initial values}\label{fig:SMVCodeInit}
\end{figure}
The movement of the UAV from one cell to the next cell is modelled by assigning one of the five neighbouring cells to the state variable \texttt{cell}. The state variables \texttt{initial\_heading} and \\
 \texttt{destination\_heading} capture the initial and destination headings of the UAV path connecting the present cell and the destination cell. The destination heading of the path becomes the initial heading for the next path and the destination cell becomes current cell, once the UAV moves to the destination cell. The SMV code modelling these transitions is shown in the Figure \ref{fig:SMVCodeNext}.
\begin{figure}[hbpt]
\centering
\begin{verbatim}
next(initial_heading) := destination_heading;
next(current_cell) := destination_cell;
\end{verbatim}
\caption{SMV code showing transitions}\label{fig:SMVCodeNext}
\end{figure}

\par
The state variables \texttt{north\_cell} and \texttt{south\_cell} model the inputs from the navigation system regarding the present position of the UAV in the search area. The values for \texttt{north\_cell} and \texttt{south\_cell} are assigned deterministically as shown in the Figure \ref{fig:SMVCodeNorth}.
\begin{figure}[hbpt]
\centering
\begin{verbatim}
north_cell := case
                current_cell[2]=1950:1;
                1:0;
              esac;
south_cell := case
                current_cell[2]=50:1;
                1:0;
              esac;
\end{verbatim}
\caption{SMV code showing assignment}\label{fig:SMVCodeNorth}
\end{figure}
The state variables like \texttt{threat\_in\_cell1} model the environment in which the UAV is operating and model the presence of threats around the UAV. As these variables are purely environmental and the UAV has no control over the location of threats, they are declared as non-deterministic. The state variables like \texttt{other\_uav\_selected\_cell1} model the decisions made by other UAVs in the group. These variables are declared as non-deterministic, as the decisions of the other UAVs are not determined by the state of UAV. As the UAV's decisions are based on the information from the sensors which senses the presence of threats in the immediate neighbourhood, only five cells around the UAV are considered as shown in the Figure \ref{fig:NeighbouringCells}. For example the presence of threat in the cell11 is modelled by assigning value 1 to the boolean state variable \texttt{threat\_in\_cell1}. The movement of the other UAVs in the group are captured by the boolean state variables. For example the selection of the any UAV in the group to move into cell1 is modelled by assigning 1 to the state variable \texttt{other\_uav\_selected\_cell1}.

\par
The destination heading is either 90$^\circ$ or 270$^\circ$ based on the present heading and the destination cell chosen. The coordinates of the destination cell depend on the selection of the neighbouring cell and the coordinates of the current cell.  A snippet of SMV code showing the assignment of destination cell coordinates to the state variable \texttt{destination\_cell} is shown in the Figure \ref{fig:SMVCodeDest}.
\begin{figure}[hbpt]
\centering
\begin{verbatim}
destination_cell:=
         case
           cell=cell1  : case
              initial_heading =90 : [current_cell[1] , current_cell[2]+100];
              initial_heading =270: [current_cell[1] , current_cell[2]-100];
                         esac;
           cell=cell2  : ..
                         ..
                         ..
               ..
               ..
         esac;
\end{verbatim}
\caption{SMV code showing assignment of destination cell coordinates}\label{fig:SMVCodeDest}
\end{figure}

\subsection{Specification and Verification of Properties}
CTL is used to express the properties that the UAV mission is expected to satisfy. The property that ``\textit{when the UAV is flying with a heading of 90$^\circ$ and not in the north-most cell, then it flies straight when there is no threat ahead and no other UAV has chosen cell1}'' is expressed by the formula:

\begin{eqnarray*}\label{spec1}
&&AG(\texttt{initial\_heading}=90 \wedge \neg\texttt{threat\_in\_cell1} \wedge \neg\texttt{other\_uav\_selected\_cell1} \wedge \\
&&\neg\texttt{north\_cell} \rightarrow \texttt{cell}=\texttt{cell1})
\end{eqnarray*}

Similarly, formula for the case when the UAV heading is 270$^\circ$ is expressed as:
\begin{eqnarray*}\label{spec1}
&&AG(\texttt{initial\_heading}=270 \wedge \neg\texttt{threat\_in\_cell1} \wedge \neg\texttt{other\_uav\_selected\_cell1} \wedge \\
&&\neg\texttt{south\_cell} \rightarrow \texttt{cell}=\texttt{cell1})
\end{eqnarray*}

The safety property that ``\textit{the UAV doesn't enter a cell when either a threat is present or other UAV in the group has already chosen to enter that cell}'' is specified by the following formula for the case of cell1:
\begin{eqnarray*}\label{spec1}
AG(\texttt{threat\_in\_cell1} \vee \texttt{other\_uav\_selected\_cell1} \rightarrow \neg (\texttt{cell}=\texttt{cell1}))
\end{eqnarray*}

The property that ``\textit{the UAV has a initial heading of either 90$^\circ$ or 270$^\circ$ in its path}'' is expressed by the following formula
\begin{eqnarray*}\label{spec1}
AG(\texttt{initial\_heading}=90 \vee \texttt{initial\_heading}=270)
\end{eqnarray*}

The Kripke model is verified against the above properties using SMV model checker. Intel 2.2 GHz processor is used for verificaion and resources used are: user time of 0.046875 sec and system time of 0.015625 sec. SMV produced a output of true for all the above properties. The situation when the UAV is deadlocked and cannot decide where to move next is found by means of violation of the property that ``\textit{the UAV chooses one of the five neighbouring cells}''. This property can be expressed in CTL as the following formula
\begin{eqnarray*}\label{spec1}
AG(\neg(\texttt{cell}=\texttt{no\_free\_cell}))
\end{eqnarray*}

\begin{figure}[h]
\centering
\includegraphics[width=15cm]{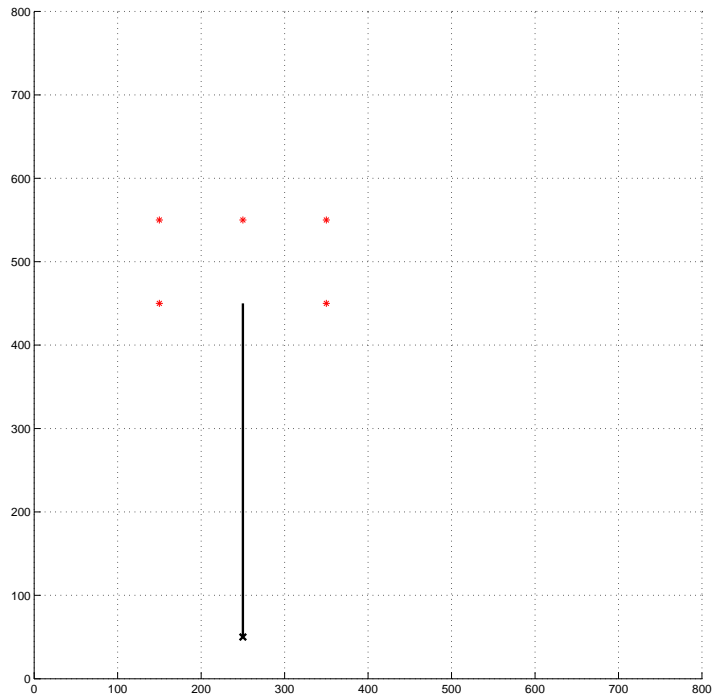}
\caption{Simulation of a case showing deadlock} \label{fig:UAVDeadlock}
\end{figure}

The violation of the above property is simulated and the Figure \ref{fig:UAVDeadlock} shows a case when the UAV is deadlocked and cannot move any further, as all the neighbouring cells contain threats.

\section{Conclusion and Future Work}
The behaviour of an UAV performing multiple-UAV cooperative search is modelled by Kripke model and some of the properties of the mission are expressed in CTL. SMV model checker is used for verifying the correctness of the model against the temporal specifications. A deadlock has been found and the trace generated by SMV has been simulated. In future, concurrency issues, properties like area coverage, liveness, reachability and fairness have to be verified.

\bibliographystyle{eptcs} 
\bibliography{Bib1}

\begin{thebibliography}{10}
\providecommand{\bibitemstart}[1]{\bibitem{#1}}
\providecommand{\bibitemend}{}
\providecommand{\bibliographystart}{}
\providecommand{\bibliographyend}{}
\providecommand{\url}[1]{\texttt{#1}}
\providecommand{\urlprefix}{Available at }
\providecommand{\bibinfo}[2]{#2}
\bibliographystart

\bibitemstart{Boskovic}
\bibinfo{author}{Jovan~D. Boskovic}, \bibinfo{author}{Ravi Prakash} \&
  \bibinfo{author}{Raman~K. Mehra} (\bibinfo{year}{2002}):
  \emph{\bibinfo{title}{A multi-layer control architecture for unmanned aerial
  vehicles}}.
\newblock {\sl \bibinfo{journal}{Proceedings of the 2002 American Control
  Conference}} , pp. \bibinfo{pages}{1825--1830}.
\bibitemend

\bibitemstart{CoordinatedMultiRobotExploration}
\bibinfo{author}{Wolfram Burgard}, \bibinfo{author}{Mark Moors},
  \bibinfo{author}{Cyrill Stachniss} \& \bibinfo{author}{Frank~E. Schneider}
  (\bibinfo{year}{2005}): \emph{\bibinfo{title}{Coordinated multi-robot
  exploration}}.
\newblock {\sl \bibinfo{journal}{IEEE Transactions on Robotics}}
  \bibinfo{volume}{21}(\bibinfo{number}{3}).
\bibitemend

\bibitemstart{Dubins}
\bibinfo{author}{L.~E. Dubins} (\bibinfo{year}{1957}): \emph{\bibinfo{title}{On
  curves of minimal length with a constraint on average curvature, and with
  prescribed initial and terminal positions and tangents}}.
\newblock {\sl \bibinfo{journal}{American Journal of Mathematics}}
  \bibinfo{volume}{79}(\bibinfo{number}{3}), pp. \bibinfo{pages}{497--516}.
\bibitemend

\bibitemstart{Clarke}
\bibinfo{author}{E.M.Clarke}, \bibinfo{author}{E.A.Emerson} \&
  \bibinfo{author}{A.P.Sistla} (\bibinfo{year}{1986}):
  \emph{\bibinfo{title}{Automatic Verification of Finite-State Concurrent
  Systems Using Temporal Logic Specifications}}.
\newblock {\sl \bibinfo{journal}{ACM Transactions on Programming Language and
  Systems}} \bibinfo{volume}{8}(\bibinfo{number}{2}), pp.
  \bibinfo{pages}{244--263}.
\bibitemend

\bibitemstart{Gat}
\bibinfo{author}{Erann Gat} (\bibinfo{year}{2004}):
  \emph{\bibinfo{title}{Autonomy Software Verification and Validation Might Not
  Be as Hard as it Seems}}.
\newblock {\sl \bibinfo{journal}{Proceedings of IEEE Aerospace Conference}} ,
  pp. \bibinfo{pages}{3123--3128}.
\bibitemend

\bibitemstart{EvolvingBehav}
\bibinfo{author}{Paolo Gaudiano}, \bibinfo{author}{Eric Bonabeau} \&
  \bibinfo{author}{Ben Shargel} (\bibinfo{year}{2005}):
  \emph{\bibinfo{title}{Evolving behaviors for a swarm of unmanned air
  vehicles}}.
\newblock {\sl \bibinfo{journal}{Proceedings of the 2005 IEEE Swarm
  Intelligence Symposium}} , pp. \bibinfo{pages}{317--324}.
\bibitemend

\bibitemstart{Hinchey}
\bibinfo{author}{Michael G.Hinchey}, \bibinfo{author}{James L.Rash} \&
  \bibinfo{author}{Christopher A.Rouff} (\bibinfo{year}{2002}):
  \emph{\bibinfo{title}{Verification and validation of autonomous systems}}.
\newblock {\sl \bibinfo{journal}{Proceedings of 26th Annual NASA Goddard
  Software Engineering Workshop}} .
\bibitemend

\bibitemstart{Brat}
\bibinfo{author}{Ari~Jonsson Guillaume~Brat} (\bibinfo{year}{2005}):
  \emph{\bibinfo{title}{Challenges in verification and validation of autonmous
  systems for space exploration}}.
\newblock {\sl \bibinfo{journal}{Proceedings of IEEE International Joint
  Conference on Neural Networks}} \bibinfo{volume}{5}, pp.
  \bibinfo{pages}{2909--2914}.
\bibitemend

\bibitemstart{ModelCheckerSPIN}
\bibinfo{author}{G.~J. Holzmann} (\bibinfo{year}{1997}):
  \emph{\bibinfo{title}{The Model Checker Spin}}.
\newblock {\sl \bibinfo{journal}{IEEE Transactions on Software Engineering}}
  \bibinfo{volume}{23}(\bibinfo{number}{5}), pp. \bibinfo{pages}{279--295}.
\bibitemend

\bibitemstart{Jayaraman1}
\bibinfo{author}{S.~Jayarman}, \bibinfo{author}{A.~Tsourdos},
  \bibinfo{author}{R.~Zbikowski} \& \bibinfo{author}{B.~White}
  (\bibinfo{year}{2006}): \emph{\bibinfo{title}{Kripke modelling approaches of
  a multiple robots systems with minimalist communication: a formal approach of
  choice}}.
\newblock {\sl \bibinfo{journal}{International Journal of System Sciences}}
  \bibinfo{volume}{37}(\bibinfo{number}{6}), pp. \bibinfo{pages}{339--349}.
\bibitemend

\bibitemstart{Kotmanova}
\bibinfo{author}{Daniela Kotmanova} (\bibinfo{year}{2008}):
  \emph{\bibinfo{title}{Temporal logic in verification of digital circuits}}.
\newblock {\sl \bibinfo{journal}{Journal of Electrical Engineering}}
  \bibinfo{volume}{59}(\bibinfo{number}{1}), pp. \bibinfo{pages}{14--21}.
\bibitemend

\bibitemstart{PowerManagement}
\bibinfo{author}{Sandeep K.Shukla} \& \bibinfo{author}{Rajesh K.Gupta}
  (\bibinfo{year}{2001}): \emph{\bibinfo{title}{A model checking approach to
  evaluating system level dynamic power management polocies for embedded
  systems}}.
\newblock {\sl \bibinfo{journal}{Proceedings of Sixth IEEE International
  High-Level Design Validation and Test Workshop}} , pp. \bibinfo{pages}{53 --
  57}.
\bibitemend

\bibitemstart{GigabitEthernet}
\bibinfo{author}{Yuan Lu} \& \bibinfo{author}{Mike Jorda}
  (\bibinfo{year}{2004}): \emph{\bibinfo{title}{Verifying a gigabit ethernet
  switch using SMV}}.
\newblock {\sl \bibinfo{journal}{Proceedings of the 41st Annual Conference on
  Design Automation}} , pp. \bibinfo{pages}{230-- 233}.
\bibitemend

\bibitemstart{ModelChecking}
\bibinfo{author}{Edmund M.Clarke}, \bibinfo{author}{Orna Grumberg} \&
  \bibinfo{author}{Doron A.Peled} (\bibinfo{year}{1999}):
  \emph{\bibinfo{title}{Model Checking}}.
\newblock {\sl \bibinfo{journal}{The MIT Press}} .
\bibitemend

\bibitemstart{McMillanSMV}
\bibinfo{author}{K.~L. McMillan} (\bibinfo{year}{1999}):
  \emph{\bibinfo{title}{Cadence. Getting started with SMV}}.
\newblock {\sl \bibinfo{journal}{Cadence Berkeley Labs}} .
\bibitemend

\bibitemstart{WebApplicationsSMV}
\bibinfo{author}{Huaikou Miao} \& \bibinfo{author}{Hongwei Zeng}
  (\bibinfo{year}{2007}): \emph{\bibinfo{title}{Model checking-based
  verification of web applications}}.
\newblock {\sl \bibinfo{journal}{Proceedings of the 12th IEEE International
  Conference on Engineering Complex Computer Systems}} , pp.
  \bibinfo{pages}{47--55}.
\bibitemend

\bibitemstart{Park}
\bibinfo{author}{Chang-Su Park}, \bibinfo{author}{Min-Jea Tahk} \&
  \bibinfo{author}{Hyochoong Bang} (\bibinfo{year}{2003}):
  \emph{\bibinfo{title}{Multiple aerial vehicles formation using swarm
  intelligence}}.
\newblock {\sl \bibinfo{journal}{Proceedings of AIAA Guidance, Navigation, and
  Control Conference and Exhibit}} .
\bibitemend

\bibitemstart{Pecheur}
\bibinfo{author}{Charles Pecheur} (\bibinfo{year}{2000}):
  \emph{\bibinfo{title}{Verification and validation of autonomy software at
  NASA}}.
\newblock {\sl \bibinfo{journal}{NASA/TM 2000-209602}} .
\bibitemend

\bibitemstart{MultirobotPlanningATimedAutomataApproach}
\bibinfo{author}{M.~M. Quottrup}, \bibinfo{author}{T.~Bak} \&
  \bibinfo{author}{R.~Izadi-Zamanabadi} (\bibinfo{year}{2004}):
  \emph{\bibinfo{title}{Multi-robot planning: a timed automata approach}}.
\newblock {\sl \bibinfo{journal}{Proceedings of the 2004 IEEE International
  Conference on Robotics and Automation}} .
\bibitemend

\bibitemstart{ChristopherRouff}
\bibinfo{author}{Christopher Rouff}, \bibinfo{author}{Mike Hinchey},
  \bibinfo{author}{Walt Truszkowski} \& \bibinfo{author}{James Rash}
  (\bibinfo{year}{2005}): \emph{\bibinfo{title}{Verifying large number of
  cooperating adaptive agents}}.
\newblock {\sl \bibinfo{journal}{Proceedings of the 11th International
  Conference on Parallel and Distributed Systems (ICPADS 2005)}}
  \bibinfo{volume}{1}, pp. \bibinfo{pages}{391--397}.
\bibitemend

\bibitemstart{Beard}
\bibinfo{author}{R.W.Beard}, \bibinfo{author}{T.W.McLain} \&
  \bibinfo{author}{D.B.Nelson} (\bibinfo{year}{2007}):
  \emph{\bibinfo{title}{Decentralized cooperative aerial surveillance using
  fixed-wing miniature UAVs}}.
\newblock {\sl \bibinfo{journal}{Proceedings of the IEEE}}
  \bibinfo{volume}{94}(\bibinfo{number}{7}), pp. \bibinfo{pages}{1306--1324}.
\bibitemend

\bibitemstart{Classification_of_dubins_sets}
\bibinfo{author}{A.~M. Shkel} \& \bibinfo{author}{V.~Lumelsky}
  (\bibinfo{year}{2001}): \emph{\bibinfo{title}{Classification of the dubins
  set}}.
\newblock {\sl \bibinfo{journal}{Robotics and Autonomous Systems}}
  \bibinfo{volume}{34}(\bibinfo{number}{4}), pp. \bibinfo{pages}{179--274}.
\bibitemend

\bibitemstart{TowardsVerifiableApproach}
\bibinfo{author}{Gopinadh Sirigineedi}, \bibinfo{author}{Antonios Tsourdos},
  \bibinfo{author}{Brian~A. White} \& \bibinfo{author}{Rafa{\l} {\.Z}bikowski}
  (\bibinfo{year}{2009}): \emph{\bibinfo{title}{Towards verifiable approach to
  mission planning for multiple UAVs}}.
\newblock {\sl \bibinfo{journal}{Proceedings of AIAA Infotech@Aerospace
  Conference and AIAA Unmanned .. Unlimited Conference}} .
\bibitemend

\bibitemstart{Coordinated_target_assignment}
\bibinfo{author}{Randal W.Beard}, \bibinfo{author}{Tinmothy W.McLain},
  \bibinfo{author}{Michael A.Goorich} \& \bibinfo{author}{Erik P.Anderson}
  (\bibinfo{year}{2002}): \emph{\bibinfo{title}{Coordinated target assignment
  and intercept for unmanned air vehicles}}.
\newblock {\sl \bibinfo{journal}{IEEE Transactions on Robotics and Automation}}
  \bibinfo{volume}{18}(\bibinfo{number}{6}), pp. \bibinfo{pages}{911--922}.
\bibitemend

\bibitemstart{Jin}
\bibinfo{author}{Y.Jin}, \bibinfo{author}{A.A.Minai} \&
  \bibinfo{author}{M.Polycarpou} (\bibinfo{year}{2003}):
  \emph{\bibinfo{title}{Cooperative real-time search and task allocation in UAV
  teams}}.
\newblock {\sl \bibinfo{journal}{Proceedings of IEEE Conference on Decision and
  Control}} , pp. \bibinfo{pages}{7--12}.
\bibitemend

\bibitemstart{Yang}
\bibinfo{author}{Y.Yang}, \bibinfo{author}{M.Polycarpou} \&
  \bibinfo{author}{A.A.Minai} (\bibinfo{year}{2007}):
  \emph{\bibinfo{title}{Multi-UAV cooperative search using an opportunistic
  learning method}}.
\newblock {\sl \bibinfo{journal}{Transactions of the ASME}}
  \bibinfo{volume}{129}, pp. \bibinfo{pages}{716--728}.
\bibitemend

\bibliographyend
\end{thebibliography}
\end{document}